\documentclass[11pt, a4paper, twocolumn, logo, copyright]{exalsius}

\pdftrailerid{redacted}

\makeatletter
\renewcommand\bibentry[1]{\nocite{#1}{\frenchspacing\@nameuse{BR@r@#1\@extra@b@citeb}}}
\makeatother

\usepackage{kantlipsum, lipsum}
\usepackage{dsfont}
\usepackage{csquotes}
\usepackage{balance}

\usepackage{todonotes}
\setuptodonotes{inline}

\usepackage{algorithm}
\usepackage{algpseudocode}
\usepackage{adjustbox}
\usepackage{multirow}
\usepackage{nicefrac}
\usepackage{graphicx}

\newcolumntype{R}[2]{%
    >{\adjustbox{angle=#1,lap=\width-(#2)}\bgroup}%
    l%
    <{\egroup}%
}

\usepackage[numbers, sort&compress]{natbib}
\usepackage{comment}

\usepackage{pifont}%

\usepackage{subcaption}
\usepackage{url}

\graphicspath{{figures/}}

\title{Distributed LLM Pretraining During Renewable Curtailment Windows: A Feasibility Study}

\codeurl{https://github.com/exalsius/curtail-llm}


\reportnumber{} %

\renewcommand{\today}{}

\author[1,3]{Philipp Wiesner}
\author[1]{Soeren Becker}
\author[2]{Brett Cornick}
\author[1]{Dominik Scheinert}
\author[1]{Alexander Acker}
\author[3]{Odej Kao}

\affil[1]{Exalsius}
\affil[2]{Deep Science Ventures}
\affil[3]{TU Berlin}

\begin{abstract}
Training large language models (LLMs) requires substantial compute and energy.
At the same time, renewable energy sources regularly produce more electricity than the grid can absorb, leading to \emph{curtailment}, the deliberate reduction of clean generation that would otherwise go to waste.
These periods represent an opportunity: if training is aligned with curtailment windows, LLMs can be pretrained using electricity that is both clean and cheap.
This technical report presents a system that performs full-parameter LLM training across geo-distributed GPU clusters during regional curtailment windows, elastically switching between local single-site training and federated multi-site synchronization as sites become available or unavailable.
Our prototype trains a 561M-parameter transformer model across three clusters using the Flower federated learning framework, with curtailment periods derived from real-world marginal carbon intensity traces.
Preliminary results show that curtailment-aware scheduling preserves training quality while reducing operational emissions to 5--12\% of single-site baselines.
\end{abstract}

\begin{document}

\maketitle

\section{Introduction}
Large language models (LLMs) have become foundational components of modern software systems, enabling applications ranging from conversational assistants to code generation and scientific workflows.
However, training state-of-the-art LLMs is resource intensive, consuming substantial GPU time and energy.
At the same time, the rapid expansion of renewable generation capacity means that grids increasingly experience periods of \emph{curtailment}: excess clean electricity that must be reduced because supply exceeds demand.
These windows of surplus renewable energy vary in timing, duration, and location, but they represent compute time that is both low-carbon and low-cost.

A growing body of work on \emph{carbon-aware computing} has studied temporal and geographic load shifting for distributed ML workloads~\cite{wiesner2024fedzero, bian2024cafe, liao2025greenfl}.
Yet, \emph{the practical feasibility of pretraining LLMs during curtailment windows remains unclear}.
Pretraining is a tightly coupled and stateful process: it requires sustained throughput to amortize fixed overheads such as compilation, data-loader warmup, and checkpoint and optimizer-state restores.
Pausing, resuming, or migrating training between regions incurs nontrivial restart and communication overhead that can dominate when curtailment windows are short or sporadic.
Recent work has shown that federated pretraining with infrequent synchronization can match or surpass centralized model quality for LLMs up to 7B parameters~\cite{sani2025photon}, suggesting that the optimization approach itself is not a bottleneck.
The open question is whether the systems overhead of elastic provisioning and communication across geo-distributed sites leaves enough useful compute time within curtailment windows.
This motivates systems that can opportunistically exploit any available curtailment window and, when multiple regions are concurrently curtailed, combine their capacity to accelerate progress.

In this technical report, we pretrain a nanochat~\cite{nanochat} d20 (561M parameters) model on 12.8B tokens across three geographically distributed GPU clusters, where site availability is governed by curtailment windows derived from real-world marginal carbon intensity traces provided by WattTime~\cite{watttime}.
Our goal is not to introduce a new learning algorithm, but to show that an operational system can support \emph{full-parameter} training, including optimizer state and multi-site synchronization, under these constraints. %

\noindent
We make the following \textbf{contributions}:
\begin{itemize}
    \item we present a reactive system design for distributed LLM pretraining during curtailment windows that supports elastic participation of training sites. The system trains locally when only a single site is curtailed, performs periodic averaging of model states when multiple regions are concurrently curtailed, and pauses when no region is curtailed.
    \item we implement the system end-to-end using the Flower~\cite{beutel2020flower} FL framework and the Exalsius real-time control plane for provisioning GPU nodes across geo-distributed clusters.
    \item we evaluate the prototype by replaying historical grid conditions during training, and report on the resulting performance, energy usage, and operational carbon emissions.
\end{itemize}

\noindent
The prototype is publicly available on Github:\\{\small \url{\the\codeurltoks}}

\section{Related Work}

\paragraph{Carbon-aware computing.}
Research in the domain of carbon-aware optimization tries to reduce the operational emissions of computing systems through scheduling.
Approaches leverage \emph{temporal shifting}, executing workloads when electricity is cleaner~\cite{Cappiello_CO2AwareAdaptationStrategies_2016, Radovanovic_Google_2022, lechowicz2023online, carbonscaler2024sigmetrics}; \emph{spatial shifting}, placing workloads where electricity is cleaner~\cite{Zheng_MitigatingCurtailment_2020, lindberg2022geographic_shifting}; and approaches that combine both dimensions~\cite{sukprasert2024eurosys, hanafy2024asplos, lechowicz2025soad, wiesner2024fedzero}.
The objective is usually to minimize the carbon intensity (in gCO\textsubscript{2}/kWh) of consumed electricity, which---depending on methodology---can vary significantly over time and locations.

\paragraph{Exploiting curtailment windows.}
A complementary line of work focuses on utilizing renewable generation that would otherwise be curtailed. Curtailment occurs when renewable supply exceeds demand, storage, and transmission capacity, forcing clean electricity to be discarded. 
California curtailed 3.4 million megawatt-hours of utility-scale wind and solar output in 2024, a 29\,\% increase from the previous year~\cite{EIA2025Curtailment}.
Prior studies on exploiting curtailment windows in computing~\cite{journal2017geo_load, SEPIA_GreenITscheduling_2018, SEPIA_NegotiationGameITEnergyManagement_2020, Zheng_MitigatingCurtailment_2020, wiesner2024fedzero} show that leveraging excess renewable supply can reduce emissions while lowering operational costs.

\paragraph{Sustainable federated learning.}
Due to their high energy demand and scheduling flexibility, AI workloads are a central target for carbon-aware optimization~\cite{dodge2022carbon_intensity_ai, xu2025green}.
In particular, federated learning enables training across geographically distributed resources~\cite{mcmahan2017fedavg} and is therefore increasingly used to exploit spatio-temporal variation in clean energy availability~\cite{liao2025greenfl, bian2024cafe, thakur2025greenfl, wiesner2024fedzero}.
For example, FedZero~\cite{wiesner2024fedzero} schedules training exclusively on excess renewable energy and idle compute capacity. However, the approach assumes fast provisioning of clients and minimal synchronization overhead, assumptions that do not hold for LLM pretraining at scale.

\paragraph{Distributed LLM pretraining.}
Recent results indicate that federated LLM pretraining can achieve performance comparable to centralized learning. DiLoCo~\cite{douillard2024diloco} demonstrates competitive convergence for distributed LLM training across 8 GPU clusters, and Photon~\cite{sani2025photon} reports perplexity matching or improving centralized baselines for models up to 7B parameters. The remaining challenge is therefore a systems problem: orchestrating elastic multi-site training under intermittent resource availability.

\section{System Design}\label{sec:design}

During pretraining, the most time and energy-intensive phase of LLM development, models are trained on large text corpora to minimize perplexity, with token budgets typically guided by scaling laws~\cite{hoffmann2022chinchilla}. This stage produces a base model for subsequent fine-tuning. 
Our goal is to execute pretraining during curtailment windows when electricity is clean and cheap.

\begin{figure*}
    \centering
    \includegraphics[width=\textwidth]{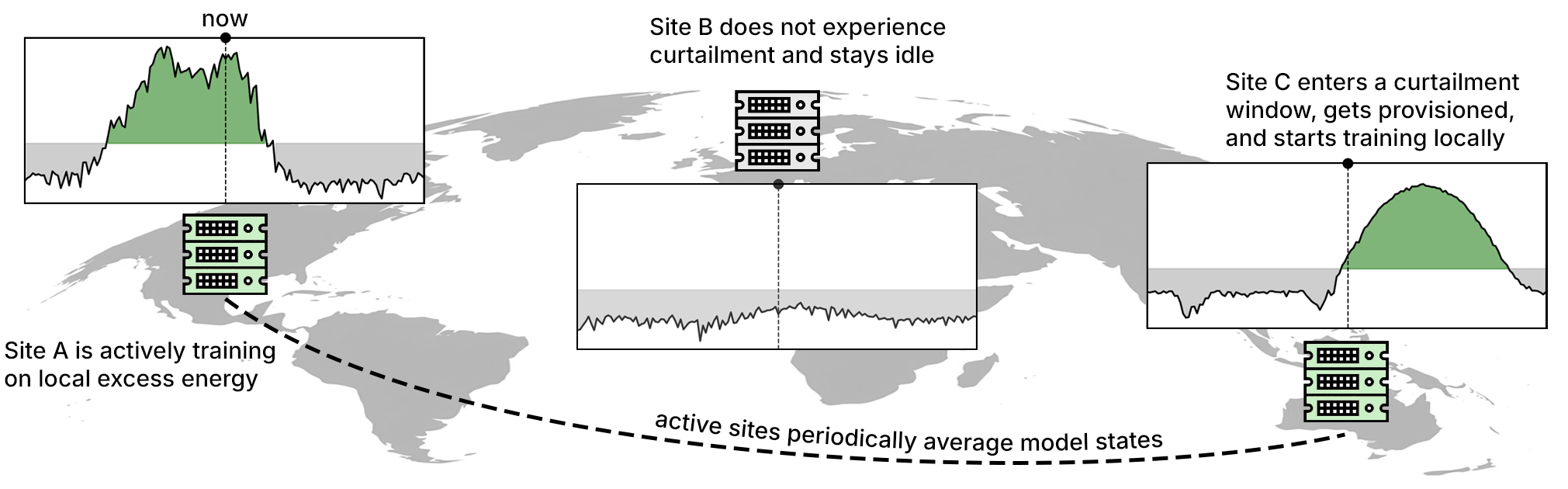}
    \caption{Sites train only during curtailment windows (green), when renewable generation exceeds demand. If multiple sites are curtailed simultaneously, they train locally in parallel and periodically average model states.}
    \label{fig:architecture}
\end{figure*}

\subsection{Problem Setting}\label{sec:design:problem}

We consider $N$ geographically distributed GPU clusters, each capable of independent local training with multi-GPU data parallelism.
An external signal indicates, for each site $s$ at time $t$, whether the local grid is currently in a curtailment window.
Let the curtailment indicator $c_s(t) \in \{0,1\}$ be~$1$ when site~$s$ is curtailed and the \emph{active set} $\mathcal{A}(t) = \{s \mid c_s(t) = 1\}$.
In practice, $c_s(t)$ can be provided by grid operators or measured directly in case of co-located renewable generation. 
Otherwise, it can be derived from drops in locational marginal prices~\cite{acun2023curtailmentprediction, maji2025curtailment}, marginal grid carbon intensity metrics~\cite{watttime}, or time-of-use renewable energy certificates~\cite{TEACS, Hourly_EAC}.

We face three challenges:

\begin{enumerate}
    \item \textbf{curtailment is sporadic and correlated}: a site may only be curtailed for short periods of time, and curtailment windows can overlap, particularly within the same balancing region. To accumulate enough compute, the system should exploit every sufficiently large window and coordinate training across sites.
    \item \textbf{curtailment is inherently uncertain}: forecasts for solar and wind carry substantial uncertainty, and curtailment additionally depends on grid congestion, demand, and operator decisions that are hard to predict. The system must therefore be fully reactive\footnote{designing for the reactive case also covers predictable schedules as a special instance.}, and aggregation must tolerate sites joining or leaving $\mathcal{A}(t)$ mid-round.
    \item \textbf{provisioning is expensive}: each provisioning event carries high cost (model transfer, optimizer warm-up, data-loader setup), and each synchronization round incurs significant communication overhead. %
\end{enumerate}

\subsection{Architecture and Execution Model}\label{sec:design:overview}

A central coordinator monitors regional curtailment signals, and provisions or deprovisions training sites. 
After provisioning, sites receive the current model $\theta$, and perform local training on their data partition.
To cope with fluctuating site availability, the system dynamically adapts its execution mode to the size of the active set $\mathcal{A}(t)$:

\begin{itemize}
  \item $|\mathcal{A}(t)| = 0$: training is suspended and no compute resources are used.
  \item $|\mathcal{A}(t)| = 1$: the active site trains continuously without synchronization, avoiding communication and orchestration overhead.
  \item $|\mathcal{A}(t)| \geq 2$: all active sites train in parallel, which we refer to as \emph{federated mode}. The coordinator initiates periodic synchronization rounds.
\end{itemize}

Figure~\ref{fig:architecture} illustrates a system state in federated mode. 
Training proceeds in timed rounds: At the start of each round, the coordinator dispatches the current global model to all active sites. Sites train locally until the round timer expires, complete their current gradient accumulation step, and return updated model parameters using work-weighted FedAvg~\cite{mcmahan2017fedavg}:
\begin{equation}\label{eq:fedavg}
  \theta =
  \sum_{s \in \mathcal{A}} 
  \frac{b_s}{\sum_{k \in \mathcal{A}} b_k}\,\theta_s,
\end{equation}
where $b_s$ denotes the number of batches processed by site $s$ during the round and $\theta_s$ its resulting model state. 
Weighting by work performed accounts for heterogeneous hardware and naturally handles sites joining or leaving during a round.

Note, that we adopt federated learning for robustness to dynamic participation and heterogeneous throughput, rather than for privacy or data locality constraints.
If data cannot be freely distributed across clients to approximate an IID distribution, the aggregation step can be replaced by more advanced methods that explicitly address data heterogeneity~\cite{scaffold2020, douillard2024diloco, grinwald2024floco, reddi2021adaptive}.

\subsection{Curtailment-Aware Provisioning}\label{sec:design:protocol}

To prevent oscillations from short signal fluctuations, we apply time-based hysteresis when reacting to the curtailment signal \(c_s(t)\).
If \(c_s(t)\) remains continuously equal to 1 for at least \(\tau_{\uparrow}\), the coordinator provisions site \(s\):
a GPU node is allocated, the training stack is deployed, and the site connects to the server.
If \(c_s(t)\) remains continuously equal to 0 for at least \(\tau_{\downarrow}\), the site is deprovisioned.
Before shutdown, the site completes its current gradient accumulation cycle, uploads model state and metrics, and terminates.
Because provisioning incurs significant overhead, we prefer to keep a site running slightly longer rather than reprovision it, and therefore typically choose \(\tau_{\downarrow} > \tau_{\uparrow}\).

\subsection{Round Sizing and Overhead}\label{sec:design:rounds}

Related work on federated LLM pretraining typically uses large local step counts (e.g., 500 steps between global aggregation~\cite{douillard2024diloco,sani2025photon}) to amortize the cost of synchronizing full model parameters.
In our prototype, rounds are instead defined by wall clock time with a configurable duration $\Delta_{\text{round}}$, rather than by local step counts. This design reduces straggler effects in heterogeneous environments or when sites join late in a round: each site trains until the timer expires and then finishes its current gradient accumulation step before stopping.

For reference:
A single aggregation round for the 561M parameter model incurs roughly 60\,s of serialization and communication overhead and about 55\,s for DDP~\cite{li2020pytorchdistributed} process setup and teardown, totaling around 115\,s of non training time per round. At $\Delta_{\text{round}} = 600$\,s, this corresponds to roughly 80\,\% compute utilization. During this interval, each client completes about 220 training steps, fewer than reported in prior work. However, those studies target models up to 7B parameters, where synchronization overheads are substantially higher. Longer rounds would further improve utilization but delay synchronization; identifying an optimal round duration is left for future work.

\subsection{Data Management}\label{sec:design:data}

The training corpus is pre-tokenized and split into $S$ fixed-size shards.
The coordinator maintains a progress vector $\mathbf{p} \in \mathbb{N}^S$, where $p_j$ records the number of rows already consumed in shard~$j$.
At the start of each round, all incomplete shards are sorted by descending progress and round-robin assigned to the active sites, so that each site receives a roughly equal partition of the remaining work. For heterogeneous setups, assignment can be weighted by expected compute capabilities.

The coordinator sends each site its assigned shard indices together with the corresponding entries of $\mathbf{p}$, allowing the site to resume each shard from the correct position.
After training, site~$s$ returns both its updated model $\theta_s$ and progress values $\{p'_j\}$ for its assigned shards.
The coordinator applies model aggregation (Equation~\ref{eq:fedavg}) and progress updates $p_j \leftarrow \max(p_j,\, p'_j)$ atomically; if a site fails to report, neither $\theta$ nor $\mathbf{p}$ are affected.

Within each multi-GPU site, row groups within the assigned shards are stride-partitioned across local DDP ranks~\cite{li2020pytorchdistributed}, ensuring that each token is consumed approximately once.

\section{Implementation}\label{sec:impl}

We prototypically implemented the system using the Exalsius~\cite{exalsius} control plane for infrastructure management and a custom Flower~\cite{beutel2020flower} Kubernetes operator for federated coordination. %

\paragraph{Provisioning and cluster integration.}

Exalsius is a multi-site resource orchestrator for heterogeneous public cloud and on-premises environments built around Kubernetes. 
It provides lifecycle management for worker nodes, automated cluster deployment and operation, and integration with workload orchestration systems. 
Dynamic resource provisioning is abstracted into a uniform resource model that allows higher-level components to reason about available capacity without provider-specific logic. 
In our prototype, GPU nodes are automatically provisioned and incorporated into clusters according to §\ref{sec:design:protocol}, while node removal triggers a graceful detachment process to ensure cluster stability.

\paragraph{Federation orchestration.}
Federated training is deployed via a custom Kubernetes operator~\cite{exalsius_flower_operator_2026} for Flower. The operator reconciles a declarative \texttt{Federation} custom resource into a centralized coordinator and multiple distributed training sites. 
Elastic membership is achieved through the Exalsius node lifecycle API. 
Newly integrated nodes automatically become eligible training sites, while deprovisioned nodes leave the federation as part of the reconciliation loop. Secure inter-site communication is provided via integrated network encryption.

\paragraph{Runtime coordination.}
The coordinator runs as a Flower \emph{SuperLink} and sites as \emph{SuperNodes}, communicating via gRPC. Each round transfers the model state along with shard assignments and metrics. A custom server-side strategy implements scheduling and round control. Round completion and graceful shutdown signals are distributed via Redis pub/sub.

\begin{figure*}[t]
    \centering
    \includegraphics[width=.95\linewidth]{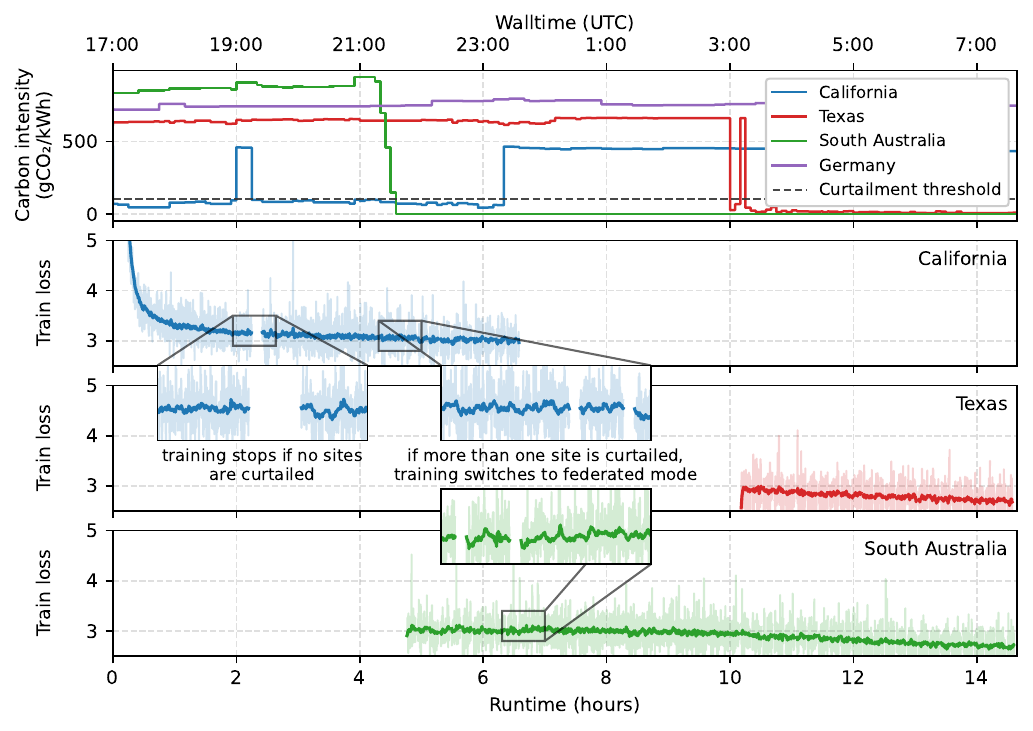}
    \vspace{-3mm}
    \caption{Execution timeline of curtailment-aware pretraining, showing how training shifts across renewable curtailment windows in California, Texas, and South Australia.}
    \label{fig:fig2}
\end{figure*}

\section{Experimental Evaluation}\label{sec:eval}

We evaluate the prototype by pretraining a small-scale LLM across geographically distributed GPU clusters based on real curtailment traces.

\paragraph{Training workload.}
We pretrain nanochat~d20, a 20-layer, 561M-parameter transformer model, following the reference implementation from the nanochat repository~\cite{nanochat}.
Training is performed on 12.8B tokens of FineWebEdu-100B~\cite{penedo2024fineweb}.

\paragraph{Infrastructure.}
We use three geographically distributed GPU clusters (sites), each equipped with four NVIDIA A100 GPUs.
Although the clusters are physically distributed, the carbon-intensity traces associated with the chosen regions do not necessarily match the physical locations of the clusters in this prototype.

\paragraph{Curtailment signal.}
We use WattTime's marginal operating emissions rate~\cite{watttime} as a proxy for renewable curtailment.
This metric estimates the emissions of the marginal generator responding to incremental demand and drops sharply when excess renewable supply displaces fossil generation.
We classify a region as curtailed when the value falls below 100 gCO\textsubscript{2}/kWh.
Historical traces starting from January 11, 2026 at 17:00 UTC are replayed with Vessim~\cite{wiesner2024vessim} to enable controlled, repeatable evaluations.

\paragraph{Region selection.}
As curtailment events can be rare and sporadic, we initially selected four regions with comparatively frequent curtailment windows: California ISO Northern, SPP North Texas, South Australia, and Germany. This selection is optimistic but allows us to stress-test elastic behavior under realistic start and stop dynamics. During the experiment period, Germany exhibited no curtailment windows, resulting in effective training across three clusters.

\subsection{Execution Timeline}

The execution timeline of the curtailment-aware execution is visualized in Figure~\ref{fig:fig2}.
We first describe significant events, starting at January 11, 2026 at 17:00 UTC.

\begin{itemize}
    \item \textbf{17:05} ---
    Carbon intensity in California drops below 100 gCO\textsubscript{2}/kWh, triggering provisioning of the first site ($\tau_{\uparrow} = 10s$). About five minutes later, the site connects to the coordinator, receives the initial model $\theta$ and progress vector~$\mathbf{p}$, and starts local training.

    \item \textbf{19:00} ---
    The curtailment window in California ends. After a 10-minute hysteresis period ($\tau_{\downarrow} = 10$\,min), the site receives a stop signal, returns the updated $\theta_s$ and $\{p'_j\}$, and is deprovisioned. Shortly after, a new curtailment window opens, the site is reprovisioned, and resumes local training.

    \item \textbf{21:40} ---
    A curtailment window opens in South Australia; the corresponding site gets provisioned and connects. Training transitions to federated mode: Every $\Delta_{\text{round}} = 10$\,min, sites receive a stop signal, return $\theta_s$ and $\{p'_j\}$, and the coordinator aggregates and redistributes weights.

    \item \textbf{23:25} ---
    The California curtailment window ends. After the hysteresis period, the site returns its progress and is deprovisioned. Training continues in South Australia.

    \item \textbf{3:20} ---
    Texas enters a curtailment window, triggering federated mode. Note that short interruptions in curtailment of $\tau_{\downarrow} = 10$\,min do not trigger deprovisioning.

    \item \textbf{8:48} ---
    Both active sites finish processing their assigned shards, and the run concludes.
\end{itemize}

\subsection{Training Performance}

\begin{figure}[b!]
    \centering
    \includegraphics[width=1\linewidth]{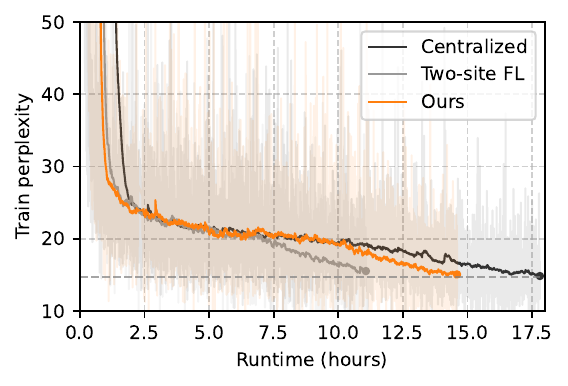}
    \vspace{-7mm}
    \caption{Curtailment-aware training reaches comparable perplexity as centralized training.}
    \label{fig:perplexity}
\end{figure}

Figure~\ref{fig:perplexity} compares training convergence under three scenarios: centralized single-site training, continuous two-site FL with $\Delta_{\text{round}} = 10$\,min, and our curtailment-aware execution.
The centralized baseline processes the full token budget in 17.8 hours and reaches a final EMA-smoothed train perplexity of 14.8.
Continuous two-site FL reduces wall-clock time to 11.1 hours but converges to a slightly higher perplexity of 15.5, reflecting the known tradeoff between parallelism and optimization noise.

Our curtailment-aware approach completes training in 14.6 hours, lying between the single- and two-site baselines in runtime.
The final EMA-smoothed perplexity is 15.1, while the best perplexity observed during training is 14.6.
Despite intermittent execution and transitions between single-site and federated modes, optimization remains stable and convergence closely matches continuous baselines.

Table~\ref{tab:runtime_energy_ppl} summarizes runtime, best achieved train perplexity, and total energy consumption.
Dynamic provisioning and synchronization introduce modest overhead, slightly increasing energy use compared to static deployments.
Nevertheless, the results show that elastic participation and intermittent availability do not materially degrade training quality.

\begin{table}
\centering
\small
\caption{Overhead from dynamic site provisioning slightly increases energy consumption.}
\vspace{-2mm}
\label{tab:runtime_energy_ppl}
\begin{tabular}{lccc}
\toprule
Scenario &
\shortstack{Runtime\\(hours)} &
\shortstack{Best train\\perplexity} &
\shortstack{Energy\\(kWh)} \\
\midrule
Centralized   & 17.8 & 14.7 & 36.0 \\
2-Site FL   & 11.1 & 15.2 & 36.1 \\
Ours          & 14.6 & 14.6 & 37.7 \\
\bottomrule
\end{tabular}
\end{table}

\subsection{Energy and Carbon Footprint}

Total energy consumption is comparable across all scenarios, ranging from 36.0 to 37.7~kWh (Table~\ref{tab:runtime_energy_ppl}).
Our approach consumes slightly more energy due to extended wall-clock time and reprovisioning overhead during gaps in curtailment availability.
Thus, emissions reductions stem from \emph{where and when} energy is consumed rather than from reduced compute.

\begin{figure}[t]
    \centering
    \includegraphics[width=1\linewidth]{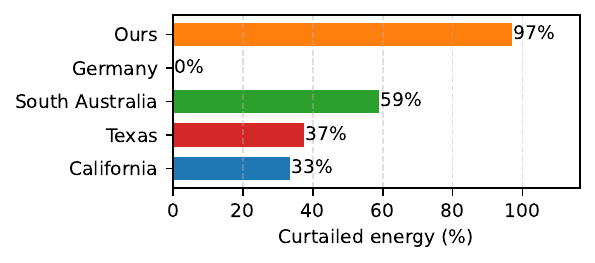}
    \vspace{-8mm}
    \caption{Fraction of training energy drawn during curtailment windows.}
    \label{fig:barchart_curtailed}
    
    \vspace{3mm}
    
    \includegraphics[width=1\linewidth]{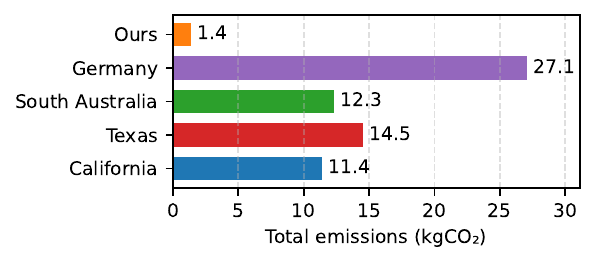}
    \vspace{-8mm}
    \caption{Operational carbon emissions computed using marginal emissions accounting~\cite{watttime}.}
    \label{fig:barchart_emissions}
\end{figure}

To demonstrate curtailment utilization, Figure~\ref{fig:barchart_curtailed} displays the fraction of energy consumed during curtailment windows.
Single-region baselines use curtailed energy only opportunistically while Germany exhibits no curtailment during the trace period.
In contrast, our scheduler shifts training almost entirely into low-carbon windows, achieving 97\% curtailed energy usage. The remaining 3\% stem from training that extends into the hysteresis period $\tau_{\downarrow}$.

We estimate operational emissions by multiplying site-level energy consumption with WattTime's marginal operating emissions rate~\cite{watttime}.
Single-region executions emit between 11.4 and 27.1~kgCO\textsubscript{2}, depending on regional grid conditions, whereas our curtailment-aware execution emits only 1.38~kgCO\textsubscript{2}.
This corresponds to approximately 5--12\% of the emissions of single-region baselines.

\section{Discussion}
\label{sec:discussion}

While our findings indicate that aligning training with curtailment windows is feasible and promising, this study also exposes limitations and open challenges that define an agenda for future work.

\paragraph{When curtailment-aware training is beneficial.}
Our evaluation targets regions with frequent curtailment, where overlapping windows enabled near continuous execution and in some cases even reduced runtime relative to a single site baseline. 
In practice, achievable throughput depends strongly on geography, seasonality, and weather conditions~\cite{maji2025curtailment}. Future work should investigate permitting limited use of \enquote{dirty} energy to provide runtime guarantees.
Beyond pretraining, curtailment-aware scheduling may benefit other energy-intensive AI workloads with flexible completion times and large compute budgets, such as reinforcement learning fine tuning and alignment, synthetic data generation, evaluation campaigns, and periodic model refreshes. 

\paragraph{Systems challenges at scale.}
The effectiveness of curtailment-aware training depends on window duration relative to provisioning, synchronization, and state restoration overheads, which can dominate short windows. 
Scaling to larger models and more sites introduces increased overheads and additional challenges such as wide-area communication costs, data movement energy overheads, and failure recovery. 
Larger fleets may smooth availability but they also increase orchestration complexity and require robust progress tracking, optimizer state consistency, and fault tolerance.
Future work could explore topology-aware synchronization~\cite{li2024networkaware} and warm standby capacity to reduce cold-start delays.

\paragraph{Carbon accounting and grid signals.}
We use marginal operating emissions rates as a proxy for curtailment, interpreting sharp drops as periods of excess clean generation. 
However, determining which signals should guide carbon-aware scheduling remains non-trivial. 
Ongoing revisions to the GHG Protocol Scope~2 Guidance~\cite{ghgprotocol2025scope2} and recent research~\cite{Gagnon2022, gorka2025ElectricityEmissions, present2024carbonfactors} highlight unresolved questions around temporal matching, marginal versus average emissions factors, and the conditions under which operational signals translate into verifiable emissions reductions. 
Identifying signals that are both operationally useful and auditable remains an important direction for future work.

\paragraph{Energy infrastructure considerations.}
At the infrastructure level, an increasing number of data centers are operated within microgrids that combine on-site or off-site generation with local battery storage~\cite{Agarwal_VirtualBattery_2021, Li_iSwitch_2012, acun2023carbon_explorer}. 
Tight integration of such energy systems with cluster scheduling could buffer intermittency and further reduce reliance on grid supply~\cite{wiesner2024vessim}. 
Moreover, strategically siting GPU clusters near renewable generation or grid bottlenecks could increase access to curtailment windows and reduce transport distances for otherwise curtailed energy~\cite{chien2019zero}. 
Depending on procurement contracts, such strategies could yield substantial cost savings for data center operators while also lowering grid congestion management costs for network operators, which in Germany alone amounted to €3.2 billion in 2023~\cite{AgoraFraunhofer2025}.

\section{Conclusion}

This technical report demonstrated that large language model pretraining can be aligned with renewable energy curtailment by elastically orchestrating geo-distributed GPU clusters. Our system dynamically switches between single-site execution and federated synchronization based on regional availability of low-carbon electricity, enabling full-parameter training under intermittent site participation. Experiments with real marginal carbon intensity traces show that curtailment-aware scheduling preserves convergence behavior and training stability while reducing operational emissions to 5--12\% of single-region baselines.

\section*{Acknowledgements}
This research was made possible through the primary support of SPRIND - Federal Agency for Breakthrough Innovation, through their investment in the development of Exalsius. We also extend our gratitude to Deep Science Ventures for their financial backing and to TU Berlin for their institutional support. Finally, we thank WattTime for providing access to their API for querying marginal operating emissions rates.

\bibliographystyle{exalsius}
\bibliography{main}

\end{document}